# An Idea of Implementing Photonic Space-Time Crystals Using Metasurfaces

ONTA SHAHRIAR[1] AND M.R.C. MAHDY[2]

*Department of Electrical and Computer Engineering, North South University, Bashundhara, Dhaka-1229, Bangladesh*
[2]*mahdy.chowdhury@northsouth.edu*

**Abstract:** Photonic space-time crystals (PSTCs) are emerging materials characterized by periodic variations in electromagnetic parameters with respect to both space and time. To date, research on PSTCs remains theoretical, with no practical realizations reported. The goal of this research was to find a way to practically realize a photonic space-time crystal and identify its potential applications. This article presents one of the very first practical realizations of PSTCs by designing and simulating a metasurface-based photonic space-time crystal using full wave simulation. We derive, calculate, and plot the dispersion relation of the metasurface-based PSTC. The working principle of the dispersion relation is discussed, along with the formation of band mode, momentum bandgap, energy bandgap, mixed energy-momentum bandgap, and second-order exceptional point eigenmodes within the dispersion relation. The characteristics of these band modes, energy bandgap, momentum bandgap, and mixed energy–momentum bandgap eigenmodes, are illustrated in this article. We demonstrate that when a transverse electric wave with an amplitude of less than 1 V/m excites the momentum bandgap or the mixed momentum bandgap eigenmode, the magnitude of the electric field component increases to more than 10 V/m. Additionally, the article details the properties of the second-order exceptional point eigenmode of the PSTC, which exists in the dispersion relation when the energy and momentum bandgaps overlap under specific conditions. It is demonstrated that when a transverse electric wave with an amplitude of less than 1 V/m excites the exceptional point eigenmode, the magnitude of the electric field component increases rapidly to more than 20 V/m. Another key achievement of our research, the application of our designed metasurface-based PSTC in creating a novel metasurface-based PSTC antenna for 6G wireless communications, is also explained in this article. The operating frequency of the antenna is 28 GHz, and it can radiate in two directions simultaneously. The absolute value of the directivity of the antenna's main lobe is 2.39, and it is directed at an angle of 35 degrees relative to the normal of the metasurface. This work will contribute to advancing the understanding of PSTCs and their potential applications.

## 1. Introduction

Photonic time crystals (PTCs) are materials whose electromagnetic parameters, such as permittivity, refractive index, etc., vary periodically with respect to time with large amplitude and high frequency, thereby exhibiting momentum bandgaps. The study of wave propagation through time-varying media began in the 1950s [1–5]. When an electromagnetic (EM) wave propagates through a PTC with a modulation frequency comparable to the EM wave's frequency, it experiences time reflection and time refraction [3, 5, 6]. This process results in two new waves with wavenumbers of opposite polarity: a time-refracted wave propagating in the same direction, and a time-reflected wave propagating in the direction opposite to the original wave due to causality constraints preventing backward time propagation. In PTCs, where electromagnetic parameters vary periodically throughout the bulk media, momentum (**P**) is conserved, resulting in the same wavenumber (**K**) for the resulting waves as the initial wave, because in electromagnetism, the momentum of the wave is, $\mathbf{P} = \hbar\mathbf{K}$. However, energy is not conserved, leading to refracted and reflected waves having different frequencies from the original wave. Periodic modulation of electromagnetic properties creates multiple time-reflected and time-refracted waves, which interfere with each other to form a PTC [2, 5, 7, 8]. This interference generates a dispersion relation for PTCs where Floquet eigenmodes are organized into a band structure with bands and bandgaps along the momentum axis. The eigenvalues of the Floquet modes $\Omega_f$ (quasi-energies) are real within the bands and complex within the momentum bandgaps. Within momentum bandgaps, the modulation of PTC's parameters can lead to exponential growth and decay of wave amplitude simultaneously, although ultimately the growth dominates. This exponentially amplifying feature of momentum bandgaps has numerous applications. Within momentum bandgaps, waves with wavenumbers falling within the bandgap range are exponentially amplified, with amplification decreasing closer to the bandgap boundaries. The formation of momentum bandgaps primarily depends on the amplitude and frequency of modulation of the PTC, influencing the breadth and height of the momentum bandgaps. Momentum bandgaps are created only when the amplitude and frequency of modulation are high. If these conditions are not satisfied, the time-reflected waves created by wave propagation through the PTC will be of low intensity, and the dispersion relation will not exhibit a wide and tall momentum bandgap. These conditions distinguish PTCs fundamentally from optical parametric amplifiers (OPAs), which operate on resonance. PTCs can be considered temporal counterparts of Photonic Crystals (PCs), where electromagnetic parameters vary periodically with respect to space instead of time. Despite their similarity in type, they have major differences: PTCs break time-translation symmetry and conserve momentum, whereas dielectric PCs break spatial-translation symmetry and conserve energy. For these reasons, the wavenumber is the preferred quantum number for analyzing PTCs, with the band structure constructed along the wavenumber axis on the dispersion relation showing momentum bands and bandgaps along that axis. PCs, on the other hand, use energy as the quantum number, depicting energy bands and bandgaps along the dispersion relation's energy axis. Another important distinction between PTCs and PCs lies in the characteristics of the eigenmodes within the bands and bandgaps. In PCs, eigenmodes within the energy bandgap exhibit exponentially decaying amplitudes due to energy conservation, which restricts growth in the energy of resulting waves. Conversely, eigenmodes within the bandgaps of PTCs exhibit both exponentially growing and decaying amplitudes, extracting energy from the modulating source. Recognizing the useful features of PTCs, researchers have heavily studied them, uncovering aspects such as momentum bandgaps [9], topological aspects [10], parity-time symmetry [11], temporal aiming [12], localization via temporal disorder [13], and others [14–18]. To practically realize PTCs that can work with electromagnetic waves having high frequencies, 3D bulk materials are needed whose electromagnetic parameters can be periodically modulated with large amplitude and high frequency, yet such materials currently do not exist. The potential for PTCs to function in the sub-terahertz and optical frequency ranges depends on epsilon-near-zero materials (ENZ) [19–23] currently under research, whose refractive index can be modulated with an amplitude around 1 and a frequency in the petahertz range. The modulation of a 3D bulk material's properties at low frequencies can be achieved using electronic components like varactors, as demonstrated by observing momentum bandgaps at microwave frequencies [8].

Modulating a material's properties practically with respect to time at low frequencies can be accomplished using electronic components such as varactors. However, implementing a large 3D network of electronic components for a 3D material is highly complex and challenging. Moreover, the electrical signals in the network's wires may emit electromagnetic waves that interfere with propagating EM waves. At optical frequencies, temporal modulation of a 3D bulk media can be achieved using laser pumping, which generates and depletes ultrafast free electrons. However, due to the focused nature of lasers, achieving uniform modulation of a material's properties throughout the bulk media is impractical. Nonetheless, modulation of a 2D surface's properties at microwave frequencies using electronic components is feasible and can be practically realized. Temporal metasurface structures have been used for frequency conversion [24] and non-reciprocal frequency generation [25]. Photonic space-time crystals involve both spatial and temporal periodicity, while temporal metasurfaces function due to only time-varying properties. Temporal metasurfaces and space-time crystals are distinct things with space-time crystals having more practical applications.

To date, only theoretical research has been conducted on Photonic Space-Time Crystals (PSTCs) [26], and practical realization remains elusive. In this article, we demonstrate our practical realization of a PSTC by designing and simulating a metasurface-based photonic space-time crystal. We explore the different eigenmodes provided by the dispersion relation of a PSTC, studying how Transverse Electric (TE) waves excite various eigenmodes while propagating along the metasurface-based PSTC. PSTCs are materials whose electromagnetic parameters are periodically modulated in both time and space. We observe that the band structure of such PSTCs contains energy bandgaps, momentum bandgaps, and novel mixed momentum–energy bandgaps due to periodic modulation in both time and space. The intriguing mixed momentum–energy bandgap is unique to PSTCs, formed by applying a modulation frequency that causes a momentum bandgap and an energy bandgap to partially overlap. The eigenmodes in these bandgaps attain both a complex Bloch wavenumber $K_B$ and a complex Floquet frequency $\Omega_f$, exhibiting coupling of energy to eigenmodes with complex quasi-energy and momentum. In the mixed momentum-energy bandgaps, the ratio between the imaginary values of the Bloch wavenumber $K_B$ and the Floquet frequency $\Omega_f$ determines the characteristics of the eigenmodes inside the mixed momentum-energy bandgap. Depending on the ratio of $\mathrm{Im}(\Omega_f)$ and $\mathrm{Im}(K_B)$, these mixed bandgaps can host different types of eigenmodes. Some mixed bandgap eigenmodes correspond to the localization of EM waves while maintaining approximately the same energy; others lead to exponential growth of EM amplitude, and some result in linear power growth of the EM wave. We found that the size of the mixed momentum–energy bandgap depends on the overlap between the energy and momentum bandgaps, and it shrinks as the overlap increases. Under strict conditions, both the energy and momentum bandgaps superimpose onto each other, causing the mixed energy–momentum bandgap to collapse to a 2nd-order exceptional point, which corresponds to an eigenmode exhibiting linear power growth. Due to the wide range of intriguing features of PSTCs, it has become important to start practically realizing PSTCs to facilitate more research and discover novel features and applications. For this purpose, in this paper, we propose a metasurface-based Photonic Space-Time Crystal (PSTC). Many studies have been conducted on metasurface-based antennas [27–30], but no one has yet been able to build a metasurface-based Photonic Space-Time Crystal (PSTC) 6G antenna. This article demonstrates how our proposed metasurface-based Photonic Space-Time Crystal (PSTC) can be applied to create an antenna for 6G wireless communications. Though practical realization of the PTCs is available in the literature [31], it is quite challenging to realize a realistic version of PSTCs. The next sections of this wok will discuss such a possible practical realization of PSTC (their notable differences with PTCs and PCs [5,7,8,31]) and their real-world possible applications.

## 2. Unitcell structure and dispersion relation of the metasurface-based photonic space-time crystal

The unit cell designed to construct the complete metasurface is depicted in Fig. 1(a) and (b). It is well known that an electric field can be created and sustained within a dielectric material between two metal plates. From this, we derived the idea to place two copper plates close to each other with a gap between them filled with air, allowing an electric field to be generated and maintained between the plates. To create an electric field, we have to apply a voltage across the copper plates, which requires connecting electrical and electronic components to the plates. To achieve this, we utilized printed circuit board (PCB) technology to place and connect the components. Two copper vias are then used to link the copper plates with the components on the PCB below them. A grounded copper layer attached to the PCB provides grounding for the electrical and electronic components placed on the printed circuit board. To isolate the copper plates from the grounded copper layer, an FR-4 dielectric is placed between the two copper plates and the grounded copper layer. This specific layout of the unitcell was selected because it offers a straightforward way to create and manipulate electric fields between the copper plates. Additionally, it is easy to fabricate and cost-effective. Another advantage of this design is that when these unitcells are combined to form a metasurface, the metasurface can generate free-space propagating waves and can also couple free-space propagating waves to a surface wave mode propagating along the metasurface. This makes it highly useful for practical applications, such as constructing a surface antenna, as demonstrated later in this paper. A detailed discussion of the full structure (and its notable differences with the PTCs [5, 7, 31]) has been discussed in the supplement article.

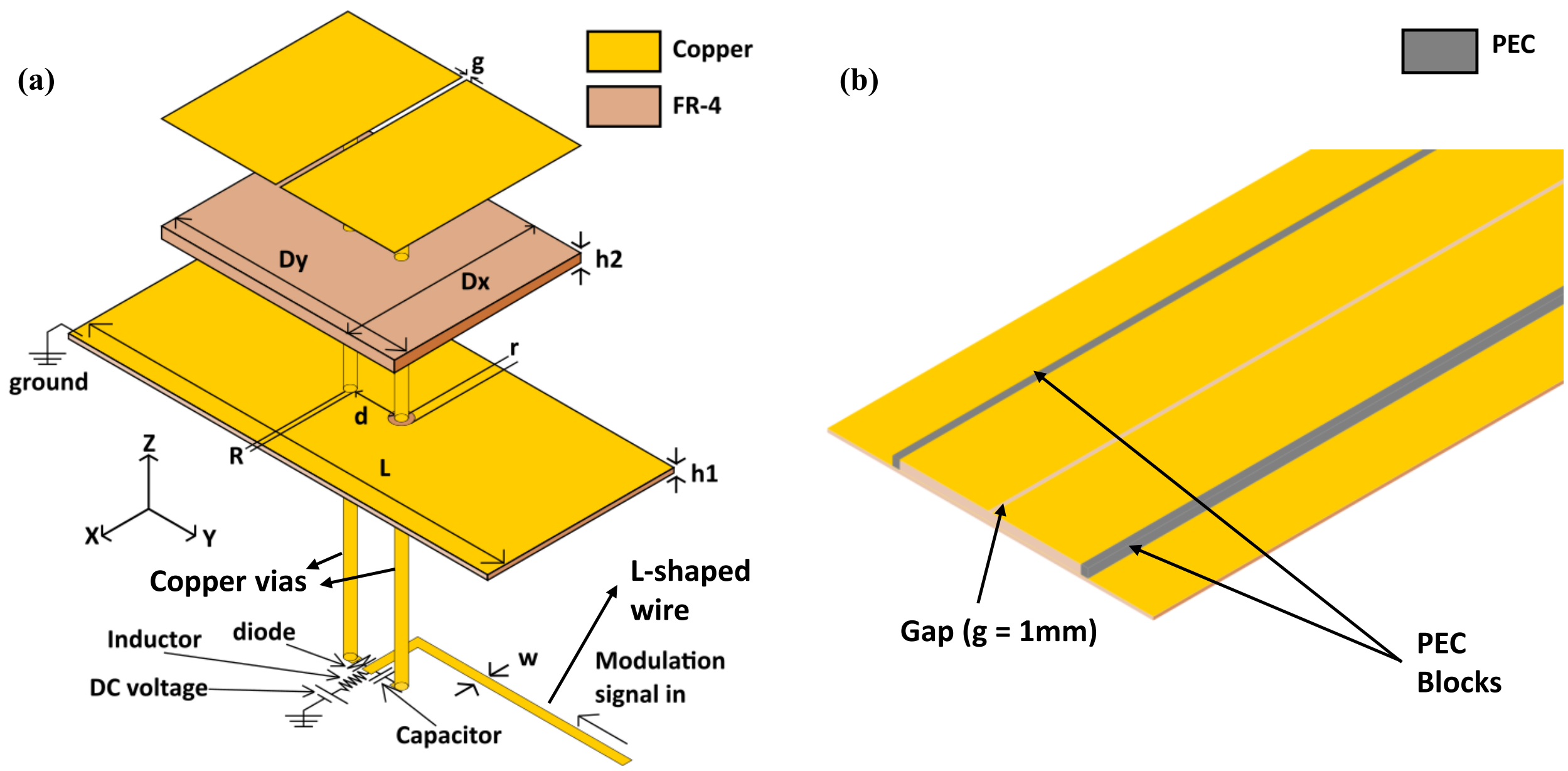

**Fig. 1.** (a) Exploded isometric view of the unitcell structure of the metasurface. g = 1mm, Dy = 50mm, Dx = 40mm, h2 = 2.4mm, h1 = 1mm, r = 2mm, R = 1mm, L = 90mm, w = 2mm, d = 9mm. (b) A partial view of the complete metasurface structure is shown here.

In CST Microwave Studio software, the structure and the circuit schematic related to the structure are designed and modeled in two different windows. The circuit schematic used to feed signals and connect circuit components to our metasurface structure in CST software is shown in the supplement article.

The sinusoidal variation of the surface capacitances of the two pairs of adjacent unitcells is 180 degrees out of phase with each other, creating a space- and time-varying capacitive surface, which functions as a photonic space-time crystal (a different concept in comparison with PTCs [5,7,31]). The complete metasurface structure, with modifications to the dimensions and dielectric material, is used to create the 6G antenna described in Section 7 of this paper. This is the structure of the metasurface-based PSTC 6G antenna we designed in this paper.

The surface capacitance of the metasurface varies simultaneously with respect to space and time. It can be written as a coordinate-separable space and time-dependent function, as shown below:

$$C(t,x) = C_0 \times C(t) \times C(x) \tag{1}$$

Here, $C_0$ is the effective surface capacitance of the unitcells used in our metasurface and is equal to 1 pF. The functions $C(t)$ and $C(x)$ are sinusoidally periodic in time and space, respectively, so they together create a space- and time-varying crystal. The surface capacitance of the metasurface-based PSTC varies according to this equation: $C(t,x) = C_0 \times (1 + (0.3 \times \cos(\omega \times t))) \times (1 + (0.3 \times \cos(k \times x)))$. In this inline equation, $\omega$ and $k$ are the angular modulation frequency ($\omega$) and spatial wavenumber ($k$) of the PSTC. Angular modulation frequency ($\omega$) is equal to $2\pi$ multiplied by the modulation frequency of the PSTC. Spatial wavenumber is equal to $2\pi$ divided by spatial period (a). The equation of our metasurface-based space-time crystal with a modulation frequency of 0.75 GHz and spatial period of 160 mm is given here:

$$C(t,x) = C_0 \times (1 + (0.3 \times \cos(4.71 \times t))) \times (1 + (0.3 \times \cos(0.04 \times x))) \tag{2}$$

The 3D figure of Eq. (2) is depicted in Fig. 3(a). Eq. (2) was plotted using the Wolfram Mathematica software. Here, we used the eigenvalue approach to find the eigenmodes. The eigenvalue approach is not only applicable to bulk media systems, it is also applicable to the metasurface structure designed by us. Here, the transverse electric surface waves propagate through the bulk air medium within the gap between the two top copper plates of the metasurface. Moreover, the modulation of surface capacitance creates alternating electric fields between the two top copper plates and these alternating fields also stay within the bulk air medium within the gap between the two top copper plates of the metasurface. So, as both transverse electric surface waves and the alternating electric fields due to changing surface capacitance stay within the bulk air medium within the gap between the two top copper plates of the metasurface, the eigenvalue approach can be used here to find the eigenmodes. We modeled the gap between the two copper plates as air by setting the relative permittivity and permeability of the background material in CST Microwave Studio to both be equal to 1. This setting fills the gap between the plates with a material whose properties match those of air. We used a coordinate-separable form of the space- and time-varying surface capacitance so that we could easily find the eigenmodes. Due to this coordinate-separable form, the electric field equation can also be written in a coordinate-separable form: $\mathbf{E} = f(t) \times g(x)\mathbf{y}$. The coordinate-separable form simplifies calculations while still maintaining the functionality of the space-time varying surface capacitance in a non-separable form. The field solution is a product of two different eigenmodes: one is the Floquet mode and the other is the Bloch mode [32, 34]. These eigenmodes can be separately found because we considered a coordinate-separable surface capacitance equation. Subsequently, these eigenmodes are used to form the dispersion relation of the PSTC, shown in Fig. 3(b). The dispersion relation is depicted only up to the first Brillouin zone [33]. The band structure comprises bands and bandgaps. Both Floquet and Bloch eigenmodes have real and imaginary parts. In band modes, both Floquet and Bloch modes have zero imaginary parts and only real parts. In energy bandgap mode, only Bloch modes have nonzero imaginary parts and a constant real part equal to 1. In momentum bandgap mode, only Floquet modes have nonzero imaginary parts and a constant real part equal to 1, as shown in Fig. 3(b). Before calculating the eigenvalues, we derive the displacement current equation for a space- and time-varying capacitive surface, as this equation will help us determine the eigenmodes. The proposed metasurface consists of multiple unitcells joined along the x-axis. Each unitcell consists of an FR-4 lossy dielectric block with two separate copper plates positioned above it, as shown in Fig. 1(a). The complete metasurface structure is depicted in the supplement article. A transverse electric (TE) wave is allowed here to propagate along the x-axis over the unitcell. As the TE wave travels, it induces positive and negative charges on the two copper plates, as depicted in Fig. 2. The polarity of the induced charge at a point on each copper plate depends on whether the electric field is directed outward from or inward towards that point. Given that the TE wave varies spatially and temporally along the unitcell, the polarity of the induced charge at a point on each copper plate also varies spatially and temporally along the x-axis over the unitcell. Therefore, the unitcell exhibits a spatially and temporally varying induced charge between its copper plates. This variation generates a corresponding spatially and temporally varying potential difference between the plates. The definition of capacitance is given in Eq. (3):

$$Capacitance(C) = \frac{Induced\ charge(Q)}{Potential\ difference(V)} \tag{3}$$

With the spatially and temporally varying induced charge and the potential difference between the two top copper plates, we achieve a corresponding spatially and temporally varying capacitive surface:

$$C(x,t) = \frac{Q(x,t)}{V(x,t)} \tag{4}$$

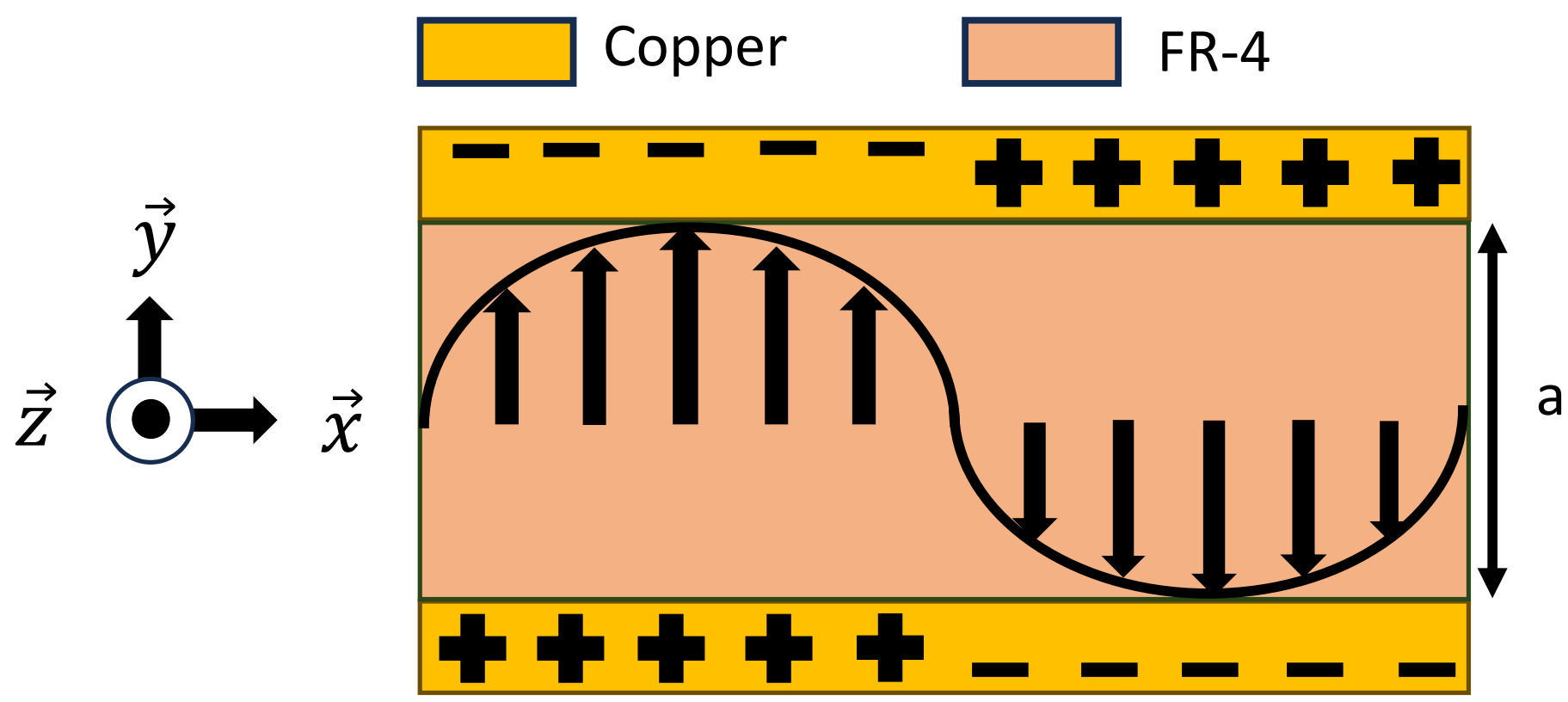

**Fig. 2.** The gap between the two top copper plates on the top of the unitcell of our proposed metasurface. Here the TE wave is shown by the black curve and arrows. The induced charges are shown as positive and negative black signs. a = 1mm, is the separation between the copper plates.

Now, we apply Ampere Maxwell's law of electromagnetism in differential form here. Here, there are no current-carrying conductors so $Ic = 0$ in Ampere Maxwell's law and only displacement current ($Id$) stays in the equation:

$$\nabla \times \mathbf{H} = Id \tag{5}$$

Ampere Maxwell's law defines the displacement current as the rate of flow of charge across the gap between the capacitor plates. So,

$$Id = \frac{\delta}{\delta \mathrm{t}} Q(x,t) \tag{6}$$

The electric field between two metal plates is equal to the potential difference between the plates divided by the distance between the plates. So, here the space and time-varying potential difference can be related to the space and time-varying electric field by the following equation,

$$V(x,t) = a \times \mathbf{E}(x,t)\mathbf{y} \tag{7}$$

By rearranging Eq. (4) and then substituting in Eq. (6), we get the following equation,

$$Id = \frac{\delta}{\delta \mathrm{t}} \big( C(x,t) \times V(x,t) \big) \tag{8}$$

Then by substituting Eq. (7) into Eq. (8) with setting a = 1 mm, we get,

$$Id = \frac{\delta}{\delta \mathrm{t}} (C(x,t) \times \mathbf{E}(x,t)\mathbf{y}) \tag{9}$$

Eq. (9) is the displacement current equation for a space and time-varying capacitive surface. Now, the calculation of eigenvalues is shown below. From Maxwell's equations for electromagnetism, we get:

$$=> \nabla \times \nabla \times \mathbf{E} = \nabla \times \left( -\frac{\delta}{\delta \mathrm{t}} \mathbf{B} \right) = -\mu_0 \times \frac{\delta}{\delta \mathrm{t}} ( \nabla \times \mathbf{H} ) \tag{10}$$

Using the Ampère-Maxwell law in differential form, along with the coordinate-separable form of surface capacitance, Eq. (1), and the displacement current equation derived above, Eq. (9), we obtain:

$$\begin{aligned} &= -\mu_0 \times C_0 \times g(x) \times C(x) \times \frac{\mathrm{d}^2}{\mathrm{dt}^2} \big( f(t) \times C(t) \big) \mathbf{y} \\ &= -\mu_0 \times C_0 \times g(x) \times C(x) \\ &\times \big( (f(\ddot{t}) \times C(t)) + (2 \times C(\dot{t}) \times f(\dot{t})) + (C(\ddot{t}) \times f(t)) \big) \mathbf{y} \end{aligned} \tag{11}$$

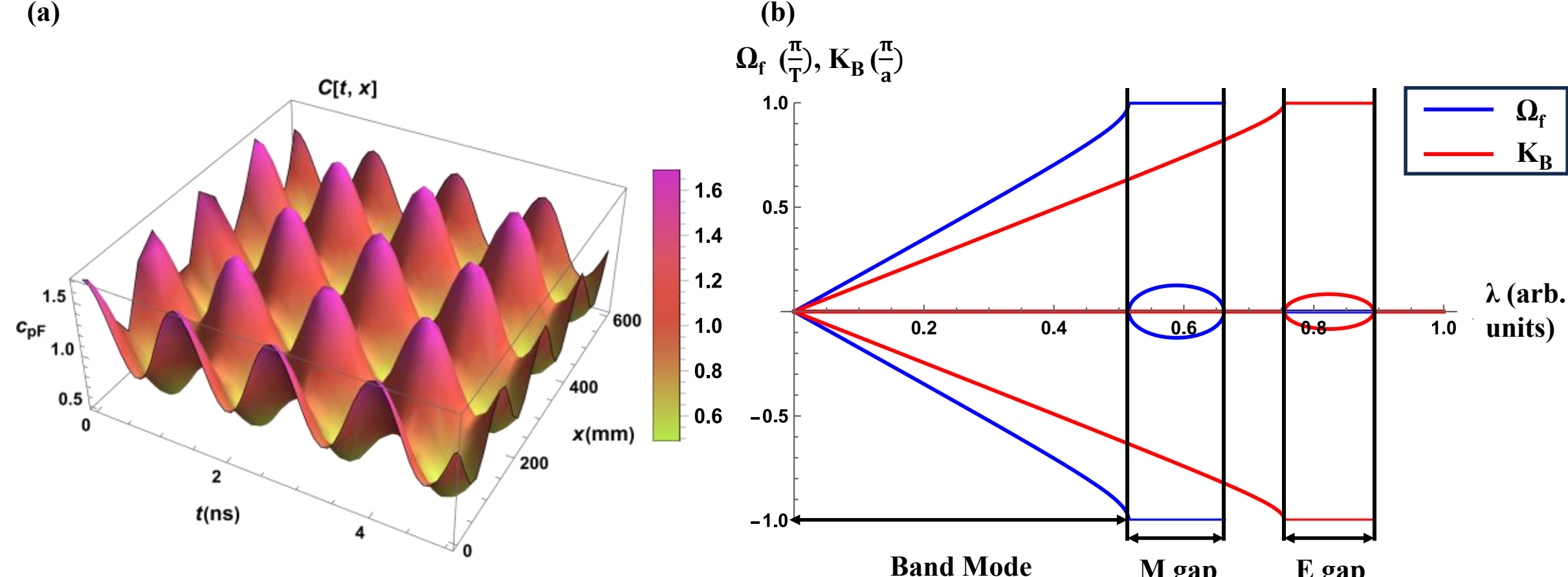

**Fig. 3.** (a) A 3D plot of the metasurface's space- and time-varying surface capacitance, created using Wolfram Mathematica software. Here, our metasurface-based PSTC crystal has a spatial period of 160 mm and a modulation frequency of 0.75 GHz. (b) The plot of the band structure of the metasurface-based photonic space-time crystal with a spatial period of 160 mm and a modulation frequency of 0.75 GHz, created using Wolfram Mathematica software. The blue curve represents the complex value of Floquet frequency ($\Omega_f$), and the red curve represents the complex value of Bloch wavenumber ($K_B$). The blue ellipse represents the imaginary part of $\Omega_f$ and the red ellipse represents the imaginary part of $K_B$. In momentum bandgap (M gap), $\Omega_f$ has a constant real part and a nonzero imaginary part. In the energy bandgap (E gap), $K_B$ has a constant real part and a nonzero imaginary part. In the Band mode, both $K_B$ and $\Omega_f$ have only real parts and no imaginary parts. λ is a unitless parameter used here for solving the coordinate-separable differential equation, Eq. (13).

Moreover, we know from Maxwell's equations for electromagnetism that,

$$=> \nabla \times \nabla \times \mathbf{E} = \nabla \cdot \nabla \cdot \mathbf{E} - \nabla^2 \mathbf{E}$$
$$= -\nabla^2\left(f(t) \times g(x)\right)\mathbf{y} \quad (12)$$

Equating Eq. (12) and Eq. (11), we get the following coordinate separable differential equation:

$$=> \frac{\frac{\mathrm{d}^2}{\mathrm{dx}^2} g(x)}{C(x) \times g(x)} = \frac{\left(\left(f(\ddot{t}) \times C(t)\right) + \left(2 \times C(\dot{t}) \times f(\dot{t})\right) + \left(C(\ddot{t}) \times f(t)\right)\right)}{\frac{f(t)}{\mu_0}}$$
$$= -\lambda^2 \quad (13)$$

As can be seen, Eq. (13) has two parts: one is a time-dependent differential equation, and the other part is a space-dependent differential equation. By solving the time-dependent differential equation for values of lambda in the range $0 \le \lambda \le 1$, We obtain the Floquet modes. By solving the space-dependent differential equation for the value of lambda in the same range $0 \le \lambda \le 1$, We obtain the Bloch modes. We solved both the time-dependent and space-dependent differential equations using the Wolfram Mathematica software. Since both $C(t)$ and $C(x)$ are periodic in their respective independent variables, the eigenmodes will take the form of the well-known Bloch wave function, which is commonly used to solve Schrödinger's wave function in periodic potential problems found in condensed matter physics [34]. The electric field equation will be the product of the Bloch and Floquet modes, as shown here,

$$\mathbf{E} = \left(\exp(ixK_B) \times \exp(it\Omega_f) \times F(x) \times F(t)\right)\mathbf{y} \quad (14)$$

Here, $K_B$ and $\Omega_f$ are the characteristic exponents of their respective Bloch wave functions, and $F(x)$ and $F(t)$ are the periodic functions of these wave functions. In Eq. (13), lambda ($\lambda$) is a unitless parameter commonly arising when solving coordinate-separable differential equations. In the context of finding Bloch modes, lambda represents angular frequency ($\omega$), while in finding Floquet modes, it represents wavenumber ($K$). By solving the time- and space-dependent differential equations of Eq. (13) for each value of lambda in the range $0 \le \lambda \le 1$, We obtain the real and imaginary parts of the characteristic exponents $K_B$ and $\Omega_f$ in Eq. (14). These are then plotted with respect to lambda using the Wolfram Mathematica software to form the dispersion relation shown in Fig. 3(b). From the band structure of the metasurface-based PSTC, we observe three types of eigenmodes. Firstly, there are band modes where both $K_B$ and $\Omega_f$ are real with no imaginary parts. Secondly, in the energy bandgap, only $K_B$ has a nonzero imaginary part and a constant real part equal to 1. Thirdly, in the momentum bandgap, only $\Omega_f$ has a nonzero imaginary part and a constant real part equal to 1. Under specific conditions, momentum and energy bandgaps may partially overlap, generating unique eigenmodes where both $K_B$ and $\Omega_f$ have nonzero imaginary parts and constant real parts equal to 1. These eigenmodes are demonstrated later in this paper. The PSTC's surface capacitance varies according to this equation:

$$C(t,x) = C_0 \times (1 + (0.3 \times \cos(\omega \times t))) \times (1 + (0.3 \times \cos(k \times x))) \quad (15)$$

Eq. (15) is a generalized version of Eq. (2), here $\omega$ and $k$ are angular modulation frequency ($\omega$) and spatial wavenumber ($k$) of the PSTC. Spatial wavenumber is equal to $2\pi$ divided by spatial period (a). Such a PSTC has multiple energy and momentum bandgaps, but we have plotted the band structure only up to the first Brillouin zone, which contains a single energy and momentum bandgap. The positions of the momentum and energy bandgaps along the $\lambda$ axis depend on the angular modulation frequency ($\omega$) and spatial wavenumber ($k$) of the PSTC respectively. Therefore, by varying $\omega$ and $k$ of the PSTC, we can change the positions of the energy and momentum bandgaps. The unique eigenmodes introduced above, arising due to the partial overlap of momentum and energy bandgaps, can be realized by increasing the modulation frequency of the PSTC from 0.75 GHz to 0.9 GHz. This will cause the two different bandgaps to partially overlap and create new eigenmodes with unique behavior.

## 3. TE wave exciting band, energy bandgap, and momentum bandgap eigenmodes

In this section, we studied TE surface wave propagation along the metasurface-based PSTC having a modulation frequency of 0.75 GHz and a spatial period of 160 mm. The dispersion relation is depicted in Fig. 3(b). In such a crystal, there are three types of eigenmodes: band mode, momentum bandgap mode, and energy bandgap mode. We excited all three types of eigenmodes and observed how the TE surface wave evolved over the metasurface-based PSTC. We used CST Microwave Studio Suite 2021 to perform FDTD simulations of TE surface wave evolution within the gap between the two top copper plates of the metasurface-based PSTC. We used the contour plot type, selected "fields on the plane" feature, and placed a cutting plane normal to the y-axis positioned at the gap situated between the two top copper plates of the metasurface in the CST software to view the contour plots of the TE wave evolution within the gap between the two top copper plates on the cutting plane. We measured and viewed the wave evolution of only the y component of the TE wave. We performed all the FDTD simulations demonstrated in this paper using CST software and the mentioned contour plot type settings. Initially, we let a TE surface wave propagate along the 2D surface, and at t = 5 ns, we activated the metasurface-based PSTC, which extends over the middle part of the metasurface. To excite the three types of eigenmodes, we launched TE surface waves with three different frequencies along the metasurface-based PSTC. Figure 4 displays different instances of wave evolution of the TE wave before the PSTC activation and after it was turned on, showing the evolution of the surface wave while propagating over the metasurface-based PSTC. From the simulation results, we found that a TE wave with a frequency of 1.5 GHz can excite the band mode. Exciting a band mode, the TE surface wave couples to the propagating modes, and the surface wave continues to propagate along the metasurface in its initial direction with the same waveform, with slight reflection, as shown in Fig. 4(a) and Fig. 4(b). We also found that a TE wave with a frequency of 1.4 GHz can excite the energy bandgap mode. In the energy bandgap mode, the surface wave stops propagating and remains in the same location over the metasurface-based PSTC while maintaining the same energy. The wave evolution can be seen in Fig. 4(c) and Fig. 4(d). Lastly, we found that a TE wave with a frequency of 3 GHz can excite a momentum bandgap mode. In momentum bandgap mode, once the PSTC is activated, the wave stops propagating, and its amplitude grows exponentially over time due to the imaginary part of the Floquet frequency ($\Omega_f$). Amplified forward and backward wave propagation then takes place along the metasurface. The surface wave evolution can be seen in Fig. 4(e) and Fig. 4(f). The 3 different eigenmodes demonstrated here using the metasurface structure can be applied in microwave photonic integrated circuits (PICs) as waveguides, amplifiers, and photonic crystals. The band mode can be used as a waveguide in PICs, the momentum bandgap eigenmode can be used as an amplifier in PICs, and the energy bandgap eigenmode can be used as a photonic crystal in PICs.

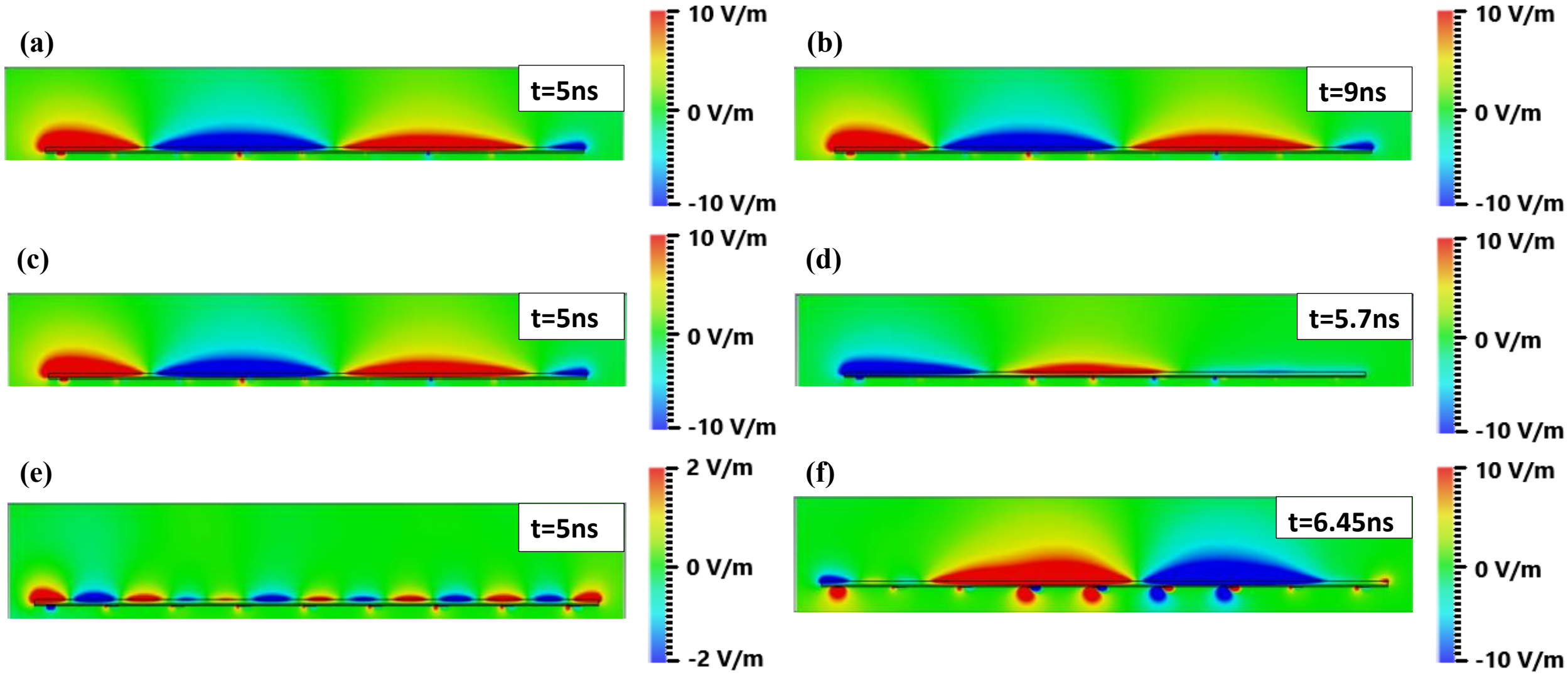

**Fig. 4.** (a) and (b) Contour plots showing wave evolution of TE wave exciting band mode. (c) and (d) Contour plots showing wave evolution of TE wave exciting energy bandgap mode. (e) and (f) Contour plots showing wave evolution of TE wave exciting momentum bandgap mode.

## 4. TE wave exciting mixed momentum-energy bandgap modes

In this section, we studied mixed momentum-energy bandgap eigenmodes. The positions of the momentum and energy bandgaps along the $\lambda$ axis in the dispersion relation of the PSTC depend on the angular modulation frequency ($\omega$) and spatial wavenumber ($k$) of the PSTC respectively. Therefore, by varying $\omega$ and $k$ of the PSTC, we can change the positions of the energy and momentum bandgaps. We found that when we increase the modulation frequency of our PSTC to 0.9 GHz from 0.75 GHz and keep the spatial period at its previous value of 160 mm, the energy and momentum bandgaps partially overlap, as can be seen in Fig. 5. This gives rise to new eigenmodes. The overlapped region is known as the lambda gap, where both Floquet and Bloch modes have nonzero imaginary parts. The lambda gap can be divided into two groups: one where Im($K_B$) is greater than Im($\Omega_f$), and in the other part, Im($\Omega_f$) is greater than Im($K_B$). Im($\Omega_f$) causes the surface TE wave's amplitude to grow exponentially over time, whereas Im($K_B$) causes the surface wave's amplitude to decay exponentially over space. In the lambda gap, Im($\Omega_f$) and Im($K_B$) oppose each other's effects. If Im($\Omega_f$) > Im($K_B$), then the exponential growth of the wave's amplitude in time dominates the nature of the eigenmode, and overall, the eigenmode resembles the momentum bandgap. If Im($\Omega_f$) < Im($K_B$), then the exponential decay of the wave's amplitude in space dominates the nature of the eigenmode, and overall, the eigenmode resembles the energy bandgap. We used CST Microwave Studio Suite 2021 to perform FDTD simulations of the TE surface wave evolution, exciting the two different eigenmodes in the lambda gap. We found that a TE wave with a frequency of 1.4 GHz can excite the mixed energy gap where Im($\Omega_f$) < Im($K_B$), and the evolution of the surface wave is shown in Fig. 6(a) and Fig. 6(b). Here, the surface wave stops propagating and becomes localized over the metasurface while maintaining the same energy, as can be seen in Fig. 6(a) and Fig. 6(b). Additionally, we found that a TE surface wave with a frequency of 3 GHz can couple to eigenmodes having Im($\Omega_f$) > Im($K_B$). In this case, the surface wave stops propagating, and its amplitude grows exponentially over time, with amplified forward and backward surface waves propagating along the metasurface. The wave evolution is shown in Fig. 6(c) and Fig. 6(d). The mixed energy bandgap eigenmode can be used in microwave photonic integrated circuits (PICs) as photonic crystals and the mixed momentum bandgap eigenmode can be used as an amplifier in PICs.

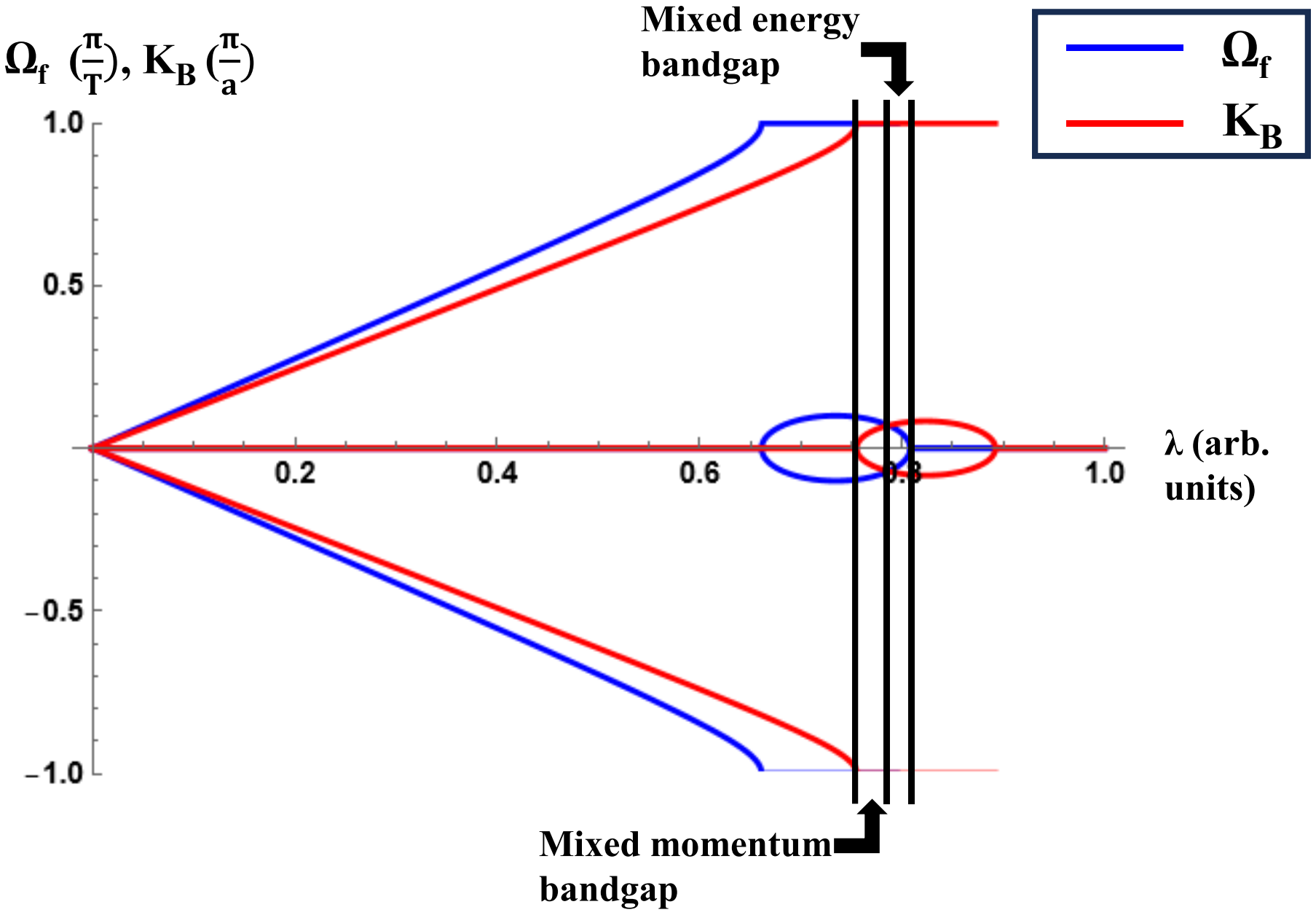

**Fig. 5.** Band structure of a metasurface-based PSTC with partially overlapping bandgaps. It was plotted using Wolfram Mathematica software. The overlapping bandgap is called the lambda gap. The lambda gap consists of the mixed energy bandgap and the mixed momentum bandgap.

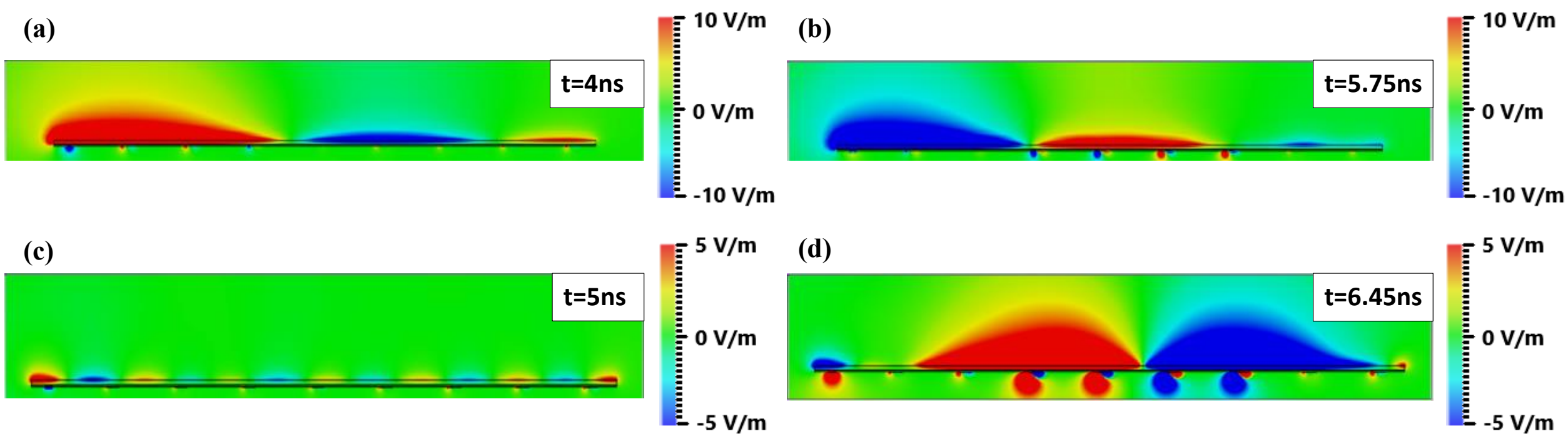

**Fig. 6.** (a) and (b) Contour plots showing wave evolution of TE surface wave exciting mixed energy bandgap mode. The PSTC is activated at t = 5ns. (c) and (d) Contour plots showing wave evolution of TE surface wave exciting mixed momentum bandgap mode.

## 5. Exceptional points of a metasurface-based photonic space-time crystal

Finally, we studied a unique case of PSTC where the energy and momentum bandgaps overlap completely, and the imaginary parts of the Floquet and Bloch modes inside the overlapping bandgaps are equal at all values of lambda within the bandgaps. The $C(t,x)$ function must satisfy two conditions for the perfect overlap of the two bandgaps. Firstly, the modulation period ($T$) of the crystal needed for the complete overlap of the momentum and energy bandgaps should be equal to the following equation:

$$T = \frac{a \times \sqrt{\max(C(t))}}{c} \tag{16}$$

Here c = speed of light in a vacuum in mm/ns, $a$ is the spatial period in mm, and max($C(t)$) is the maximum value of time-dependent surface capacitance in pF. Here, our PSTC has $a$ = 160mm and max(($C(t)$) = 1.3pF. Plugging these values into Eq. (16), we get $T$ = 0.61ns. Then, we calculated the angular modulation frequency ($\omega$) of the PSTC using the Eq. (17) and $T$ = 0.61ns,

$$\omega = \frac{2\times\pi}{T} \tag{17}$$

Then, we formed the new $C(t)$ equation using the new $\omega$, and we used the previous $C(x)$ equation because we did not change the spatial period ($a$). The new equations of $C(t)$ and $C(x)$ are given below:

$$C(t) = (1 + (0.3 \times \cos(10.33 \times t))) \tag{18}$$

$$C(x) = (1 + (0.3 \times \cos(0.04 \times x))) \tag{19}$$

For $0 \leq x \leq a$ and $0 \leq t \leq T$, where $a$ = 160mm and $T$ = 0.61ns, let

$$\beta = \frac{x}{a} = \frac{t}{T}\ ; 0 \leq \beta \leq 1 \tag{20}$$

The product of the time-dependent surface capacitance equation, Eq. (18), and the space-dependent surface capacitance equation, Eq. (19), at any value of beta($\beta$), has to be equal to a real positive constant number (L).

$$C_t(\beta) \times C_x(\beta) = \text{L} \tag{21}$$

This is the second condition, and we found that our crystal, with $a$ = 160 mm and $T$ = 0.61 ns, satisfies this condition with L = 1.69 for any value of beta ($\beta$) in the range: $0 \leq \beta \leq 1$. Exceptional points are formed by the coalescence of eigenvalues and their corresponding eigenvectors. It is known that the edges of both the energy bandgap and momentum bandgap consist of exceptional points [34]. Similarly, in our case, the bandgap edges in the dispersion relation of the PSTC also contain exceptional points. Therefore, when the energy and momentum bandgaps overlap perfectly, the exceptional points and corresponding eigenvalues also overlap, forming 2nd-order exceptional points. This overlapping of exceptional points to form higher-order exceptional points is unique to this system and produces eigenmodes with unique behavior. Here, we performed FDTD simulations of the evolution of a TE surface wave coupling to the 2nd-order exceptional point. When we set the modulation frequency of the PSTC to 1.64 GHz, we found that our metasurface-based PSTC has the dispersion relation shown in Fig. 7. We observed that a TE surface wave with a frequency of 3 GHz can couple to this 2nd-order exceptional point. The evolution of the wave is shown in Fig. 8. It can be seen from the figure that the initially propagating surface wave gets instantaneously amplified when the PSTC is turned on. The amplified surface wave continues propagating in the same direction, and an amplified surface wave is also backward radiated in the opposite direction.

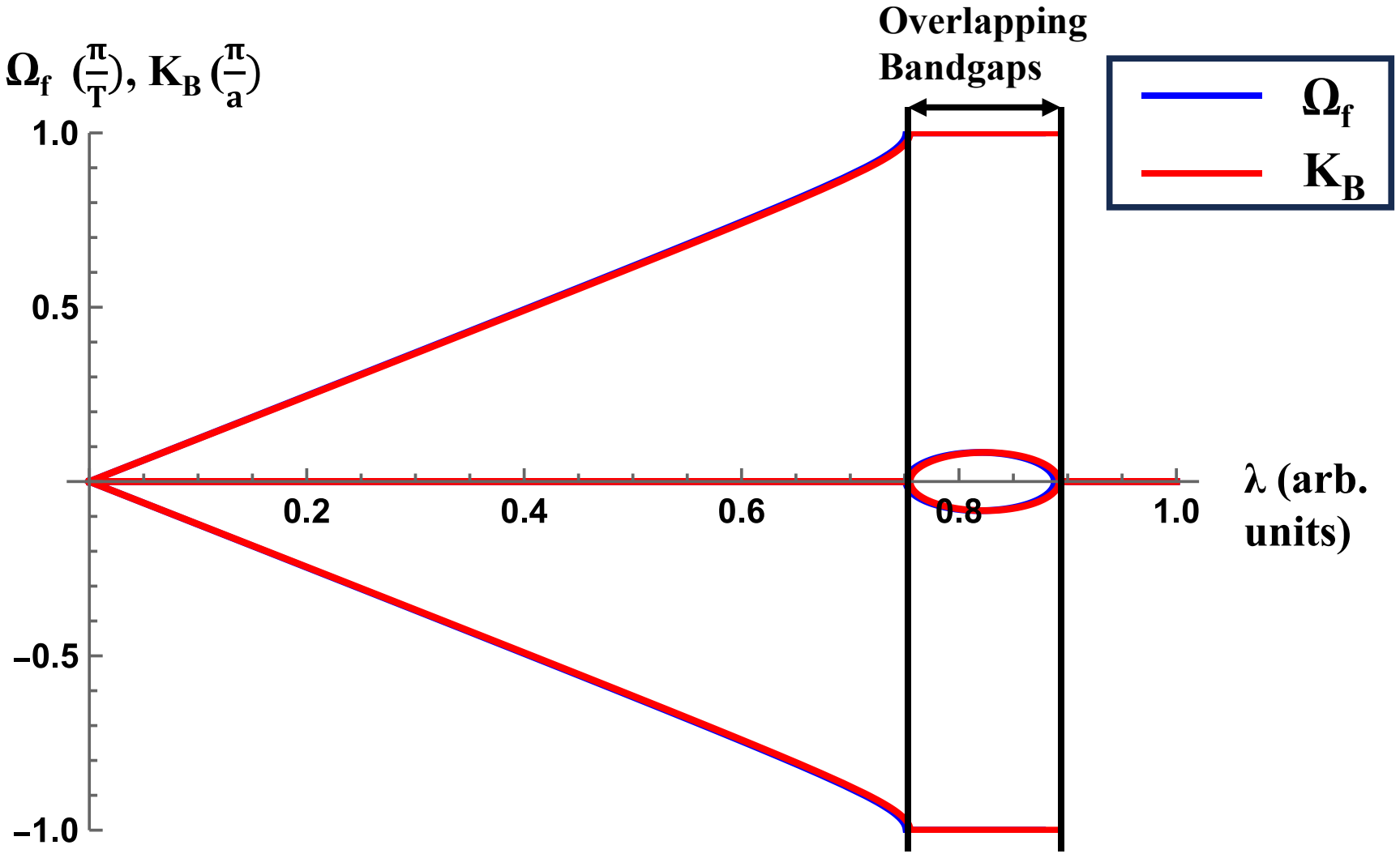

**Fig. 7.** Dispersion relation of a metasurface-based PSTC showing perfect overlap of the energy and momentum bandgaps. It was plotted using Wolfram Mathematica software.

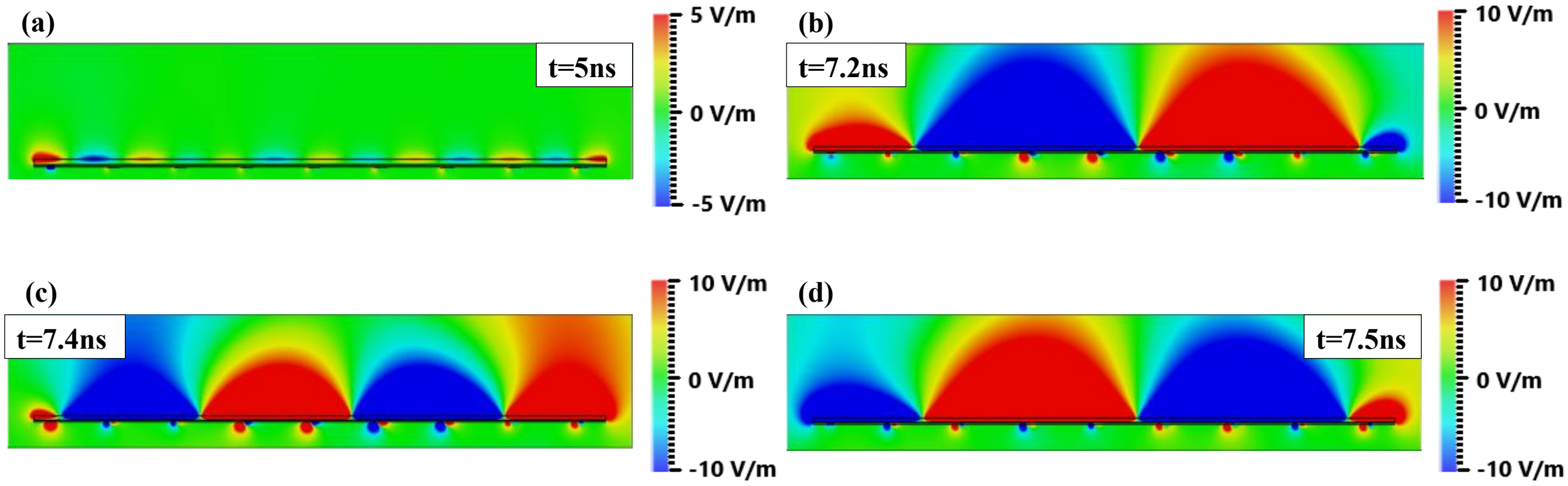

**Fig. 8.** (a) Contour plot showing TE surface wave before the PSTC is turned on at t = 5 ns. The lobes of $\overrightarrow{E_y}$ can barely be seen in Figure (a) because it has a low amplitude. (b), (c), and (d) Contour plots showing wave evolution of TE surface wave after the PSTC is turned on at t = 5ns.

## 6. Free-space wave excitation of a metasurface-based photonic space-time crystal

In this section, we illustrate that a two-dimensional metasurface-based photonic space-time crystal can be stimulated by both TE surface waves and free-space TE waves falling on the metasurface at an angle. This capability allows the metasurface to function as an antenna. Various eigenmodes such as band mode, momentum bandgap mode, energy bandgap mode, and others of the metasurface-based photonic space-time crystal can be excited by free-space wave excitation. For demonstration purposes, we launched a TE free-space wave incident on the metasurface at a 40-degree angle using a waveguide antenna in CST FDTD simulations to illustrate the free-space wave excitation of the exceptional point eigenmode of the metasurface-based PSTC. The frequency of the TE free-space wave was 3 GHz, the modulation frequency of the metasurface-based PSTC was set to 1.64 GHz, and the spatial period was set to $a$ = 160 mm. Snapshots of the wave evolution are shown in Fig. 9. In Fig. 9(a), it can be observed that the free-space TE wave couples to TE surface wave propagating modes along the metasurface, with a small reflection of the free-space wave visible over the metasurface before the PSTC is activated. At t = 5 ns, the waveguide antenna is turned off, ceasing further radiation of free-space TE waves. It is evident in Figs. 9(b), 9(c), and 9(d), that after the PSTC is activated at t = 5 ns, wave propagation stops, the wave amplitude rapidly amplifies, and the amplified surface wave continues propagating in its original direction. Additionally, an amplified surface wave is radiated in the opposite direction.

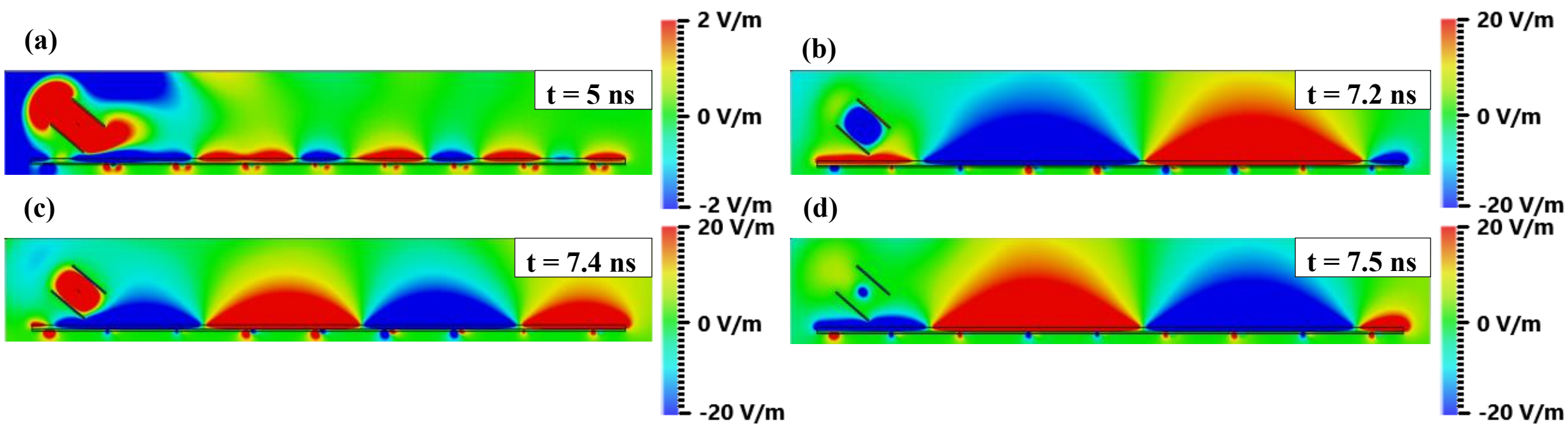

**Fig. 9.** (a) Contour plot showing TE free-space wave getting coupled to TE surface wave mode with a slight reflection of the metasurface. The reflection of the free-space wave increases with the angle made by the incident wave with the surface. ((b), (c), and (d)) Contour plots showing wave evolution of the TE surface wave exciting the exceptional point mode after the PSTC is turned on at t = 5ns.

## 7. Possibility of making an antenna for 6G wireless communications using metasurface-based PSTC

Our proposed metasurface-based PSTC can be used as an antenna, capable of working with both surface and free-space waves, as demonstrated in the previous section. Here, we have shown that by modifying the metasurface used in the previous all other simulations, a metasurface-based PSTC can be designed to operate at the high frequencies used in 6G communications. We replaced the lossy FR-4 dielectric, used in the original metasurface structure, with a lossy aluminum nitride dielectric, which has a lower loss tangent (tan δ) value of 0.0003, permittivity (ε) = 8.6 F/m, and permeability (μ) = 1 H/m. Additionally, we scaled down the metasurface structure by a factor of 0.067 to improve the transmission of TE surface waves at high frequencies and conducted S-parameter simulations of the modified PSTC structure using CST Microwave Studio software. The results of these simulations are depicted in Fig. 10.

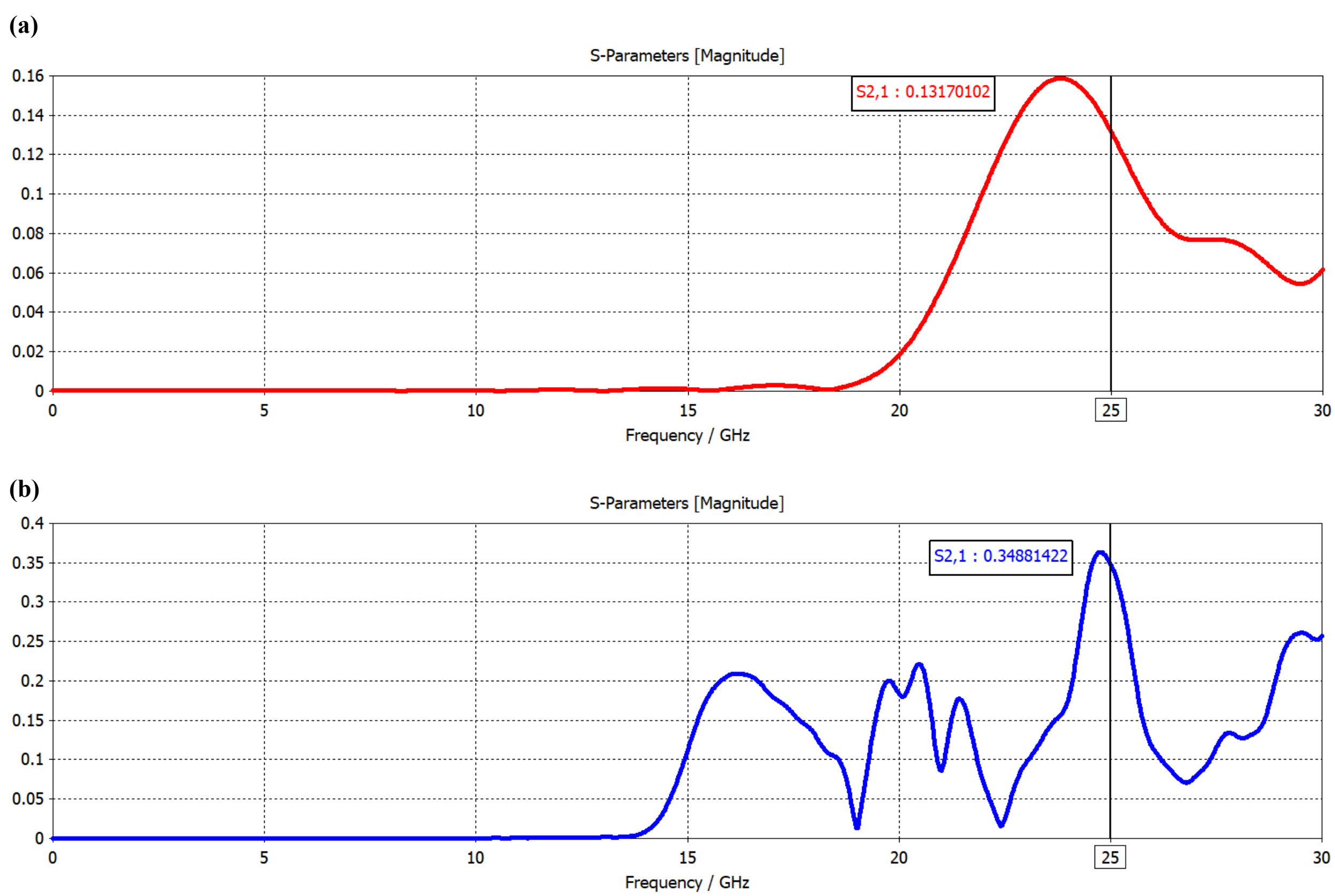

**Fig. 10.** (a) Plot of the transmission coefficient (S2,1) of the metasurface structure with lossy FR-4 dielectric. (b) Plot of the (S2,1) for metasurface structure with lossy aluminum nitride dielectric.

The S-parameter results shown in Fig. 10 indicate that the modified metasurface structure using lossy aluminum nitride as the dielectric allows TE surface waves to propagate along the metasurface or free-space waves to couple surface wave propagating modes having frequencies in the frequency range between 25GHz to 30GHz. The transmission coefficient is maximum in the frequency range of 25GHz to 30GHz. The transmission coefficient (S2,1) represents the fraction of the total power in the surface wave coming from the lefthand side of the metasurface that can reach the righthand side of the metasurface. In Fig. 10, we observe that for the lossy FR-4 dielectric, the (S2,1) value is 0.13 at 25 GHz, whereas for the lossy aluminum nitride dielectric, the (S2,1) value is 0.35 at 25 GHz. The difference in the S-parameter results of the metasurface with these two different dielectrics is due to the loss tangent (*tan δ*) value of each dielectric. The loss tangent of any lossy dielectric is the ratio of the conduction current density to the displacement current density.

$$Loss\ Tangent, \tan\delta = \frac{J_c}{J_D} = \frac{\sigma}{\omega\varepsilon} = \frac{\sigma}{2\times\pi\times f\times\varepsilon} \qquad (22)$$

Here, $\sigma$ represents the electrical conductivity of the dielectric material, $\varepsilon$ is the dielectric permittivity, and $f$ is the frequency of the EM wave. Lossy aluminum nitride has a loss tangent value equal to 0.0003, whereas lossy FR-4 has a loss tangent value of 0.025. Therefore, due to FR-4 having a larger loss tangent value compared to aluminum nitride, a metasurface with aluminum nitride as the dielectric allows EM waves at high frequencies to propagate with low power loss. Aluminum nitride acts more as an insulator than a conductor due to its low ratio between conduction current density and displacement current density. Therefore, we can utilize this scaled-down metasurface-based PSTC with lossy aluminum nitride dielectric as a 5G and 6G antenna for wireless communications, as this frequency range corresponds to the 5G high-band and 6G upper mid-band frequencies which are under consideration for 6G. To validate this, we conducted FDTD simulations of a TE surface wave with a frequency of 25GHz propagating along the metasurface. At this frequency, the transmission coefficient is maximum, as shown in the S-parameter results. We found that with a crystal having a modulation frequency of 28GHz and a spatial period of 10.7mm, the TE surface wave at 25GHz frequency can excite an exceptional point mode. The wave evolution is depicted in Fig. 11. When the PSTC is activated at t = 2.5ns, the TE wave ceases propagation, and its amplitude is instantaneously amplified. The amplified surface wave continues propagating in its original direction, while an amplified surface wave is also radiated in the opposite direction. The metasurface also radiates into free space, as shown in Fig. 11. Hence, the metasurface-based PSTC serves as an antenna radiating waves, as illustrated in Fig. 11(b), Fig. 11(c), and Fig. 11(d).

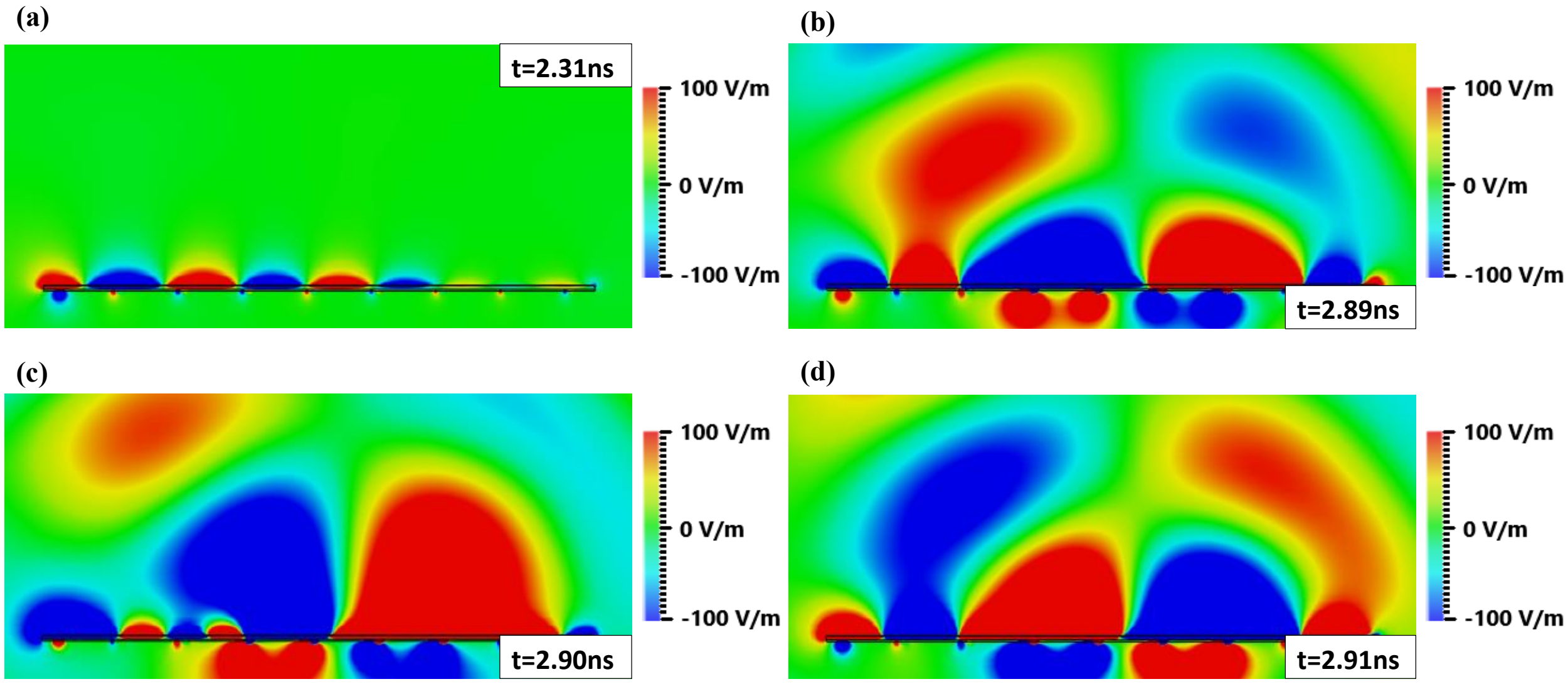

**Fig. 11.** (a) Contour plot showing TE surface wave propagating along the metasurface before the PSTC is turned on at t = 2.5ns. (b), (c), and (d) Contour plots showing TE surface wave getting coupled to an exceptional point eigenmode. The surface wave stops propagating, the surface wave's amplitude rapidly amplifies, the amplified surface wave keeps propagating along the initial direction, and an amplified surface wave is also radiated in the opposite direction. Free space radiation also happens as can be seen in (b), (c), and (d).

Some of the antenna characteristics of the 6G antenna discussed above are mentioned here. The Radar Cross Section (RCS) plot of the metasurface-based photonic space-time crystal-based antenna with a modulation frequency of 28 GHz is depicted in Fig. 12. This RCS plot was plotted using CST Microwave Studio software.

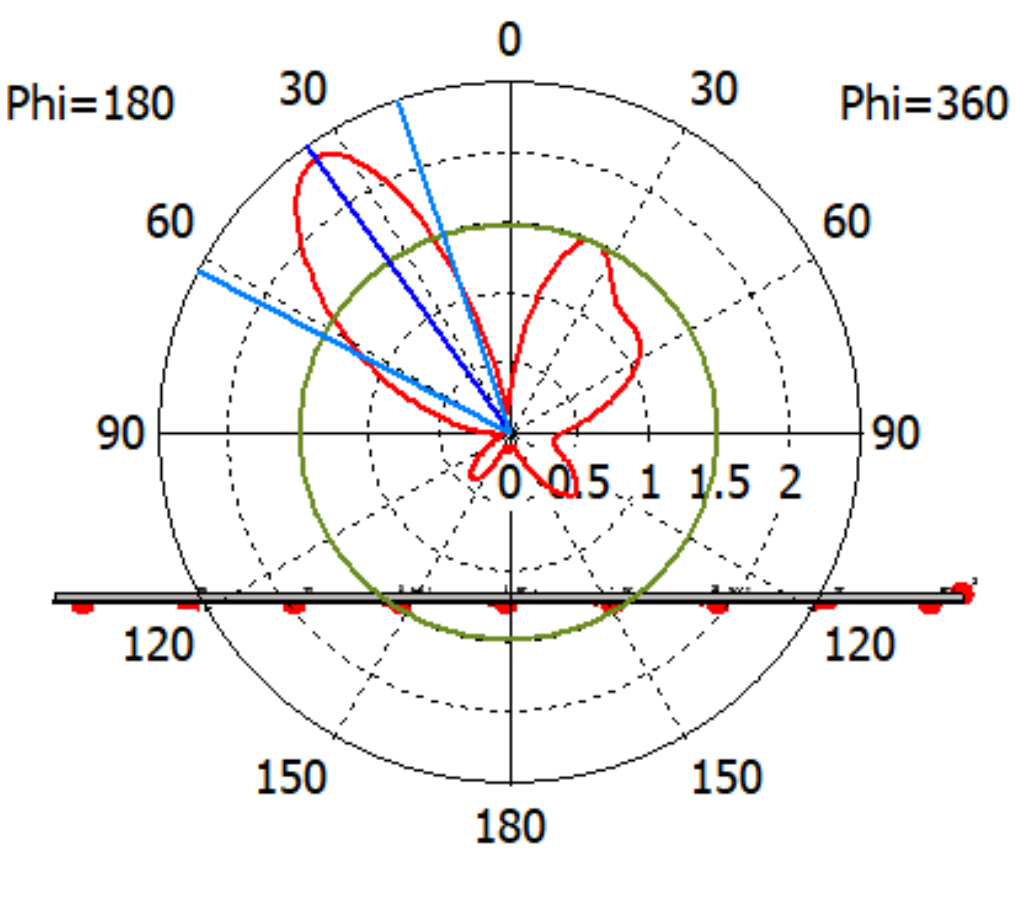

**Fig. 12.** RCS plot of the antenna showing 2 lobes. This antenna can radiate in 2 directions simultaneously.

The directivity of the main lobe is 2.39 and it is directed at an angle of 35 degrees relative to the normal to the metasurface, as shown in Fig. 12. The antenna's main lobe has a beam width of 43.6 degrees, and the side lobe level is -2.0 dB. This antenna can be utilized in antenna array systems such as reconfigurable intelligent surfaces (RIS) [36]. This antenna is a novel metasurface-based photonic space-time crystal-based 6G antenna, and its current performance metrics are acceptable. However, future work can further improve the antenna's performance.

## 8. Conclusion

In conclusion, this paper proposes, designs, and simulates for the first time a metasurface-based photonic space-time crystal, a 2D surface whose surface capacitance varies with respect to time and space simultaneously. We derive and plot the dispersion relation of the metasurface-based PSTC. The working principle of the dispersion relation is discussed along with how the mixed energy bandgap, mixed momentum bandgap, and the 2nd-order exceptional point eigenmodes form in the dispersion relation. We demonstrate the characteristic behavior of different eigenmodes corresponding to the band and various types of bandgaps in the dispersion relation by simulating them in the CST studio software. The paper showcases the capability of the momentum bandgap and the mixed momentum bandgap eigenmodes of the metasurface-based PSTC to amplify the magnitude of the electric field component of the TE wave to more than 10 V/m from below 1 V/m. Additionally, the article demonstrates the TE wave evolution of the 2nd-order exceptional point eigenmodes that exist in the dispersion relation when the energy and momentum bandgaps completely overlap in the dispersion relation of the metasurface-based PSTC. The exceptional point eigenmode can rapidly amplify the magnitude of the electric field component of the TE wave from below 1 V/m to more than 20 V/m. Furthermore, we show that all eigenmodes of our designed metasurface-based PSTC can be excited by free-space TE waves, making it suitable for use as an antenna. We design and simulate a PSTC working not only in the radio frequency domain but also in 5G high-band and 6G upper mid-band frequency ranges which are currently under consideration for 6G use cases. By changing the substrate material of the metasurface-based PSTC and scaling down its structure by a factor of 0.067, we verify in this article that our metasurface-based PSTC can be utilized as an antenna for 5G/6G wireless communications operating at 28 GHz frequency. The antenna can simultaneously emit radiation in two directions, each making an angle of 35 degrees with the normal to the metasurface structure. The two lobes of the antenna have a beam width of 43.6 degrees, enabling it to supply a wide area with its signal. All of the simulation results verify the theoretical work done in this article. The realization of photonic space-time crystals (PSTC) using metasurfaces, demonstrated in this paper, will aid researchers in further exploring PSTCs, advancing their understanding of PSTCs, and finding new applications of PSTCs.

**Funding.** M. R. C. Mahdy acknowledges the support of the internal CTRGC grant 2023-24 of North South University (approved by the members of BOT), Bangladesh.

**Disclosures.** The authors declare no conflicts of interest.

**Data availability.** The data that support the findings of this study are available from the corresponding author upon reasonable request.

**Supplemental document.** See Supplement for supporting content.

## References

1. F. R. Morgenthaler, "Velocity Modulation of Electromagnetic Waves," in *IRE Transactions on Microwave Theory and Techniques*, vol. 6, no. 2, pp. 167-172, April 1958.
2. D. Holberg and K. Kunz, "Parametric properties of fields in a slab of time-varying permittivity," in *IEEE Transactions on Antennas and Propagation*, vol. 14, no. 2, pp. 183-194, March 1966.
3. J. T. Mendonça, A. Guerreiro, and A. M. Martins, "Quantum theory of time refraction," in *Phys. Rev. A* 62, 033805, 2000.
4. A. B. Shvartsburg, "Optics of nonstationary media," in *Phys. Usp.* 48, 797, 2005.
5. F. Biancalana, A. Amann, A. V. Uskov, et al., "Dynamics of light propagation in spatiotemporal dielectric structures," in *Phys. Rev. E* 75, 046607, 2007.
6. J. T. Mendonça and P. K. Shukla, "Time refraction and time reflection: two basic concepts," in *Phys. Scripta* 65, 160, 2002.
7. J. R. Zurita-Sánchez, J. H. Abundis-Patiño, and P. Halevi, "Pulse propagation through a slab with time-periodic dielectric function ε(t)," in *Opt. Express* 20, 5586–5600, 2012.
8. J. R. Reyes-Ayona and P. Halevi, "Observation of genuine wave vector (k or β) gap in a dynamic transmission line and temporal photonic crystals," in *Appl. Phys. Lett.* 107, 074101, 2015.
9. A. M. Shaltout, J. Fang, A. V. Kildishev, et al., "Photonic time-crystals and momentum band-gaps," in *Conf. Lasers Electro-Optics*, 2016.
10. E. Lustig, Y. Sharabi, and M. Segev, "Topological aspects of photonic time crystals," in *Optica* 5, 1390–1395, 2018.
11. N. Wang, Z.-Q. Zhang, and C. T. Chan, "Photonic Floquet media with a complex time-periodic permittivity," in *Phys. Rev. B* 98, 085142, 2018.
12. V. Pacheco-Peña and N. Engheta, "Temporal aiming," in *Light Sci. Appl.* 9, 129, 2020.
13. Y. Sharabi, E. Lustig, and M. Segev, "Disordered photonic time crystals," in *Phys. Rev. Lett.* 126, 163902, 2021.
14. G. Castaldi, V. Pacheco-Peña, M. Moccia, et al., "Exploiting space-time duality in the synthesis of impedance transformers via temporal metamaterials," in *Nanophotonics* 10, 3687–3699, 2021.
15. F. Ding, A. Pors, S. I. Bozhevolnyi, et al., "Spatiotemporal isotropic-to-anisotropic meta-atoms," in *New J. Phys.* 23, 095006, 2021.
16. H. Li, S. Yin, E. Galiffi, et al., "Temporal parity-time symmetry for extreme energy transformations," in *Phys. Rev. Lett.* 127, 153903, 2021.
17. D. Ramaccia, A. Alu, A. Toscano, et al., "Propagation and scattering effects in metastructures based on temporal metamaterials," in *15th Int. Congr. Artif. Mater. Nov. Wave Phenomena, Metamaterials*, pp. 356–358, 2021.
18. R. Maas, J. Parsons, N. Engheta, et al., "Experimental realization of an epsilon-near-zero metamaterial at visible wavelengths," in *Nat. Photonics* 7, 907–912, 2013.
19. L. Caspani, R. P. M. Kaipurath, M. Clerici, et al., "Enhanced nonlinear refractive index in ε-near-zero materials," in *Phys. Rev. Lett.* 116, 233901, 2016.
20. D. M. Solis and N. Engheta, "A theoretical explanation for enhanced nonlinear response in epsilon-near-zero media," in *CLEO*, 2020.
21. Y. Zhou, M. Z. Alam, M. Karimi, et al., "Broadband frequency translation through time refraction in an epsilon-near-zero material," in *Nat. Commun.* 11, 2180, 2020.
22. V. Bruno, C. Devault, S. Vezzoli, et al., "Negative refraction in time-varying strongly coupled plasmonic-antenna-epsilon-near-zero systems," in *Phys. Rev. Lett.* 124, 043902, 2020.
23. V. Bruno, S. Vezzoli, C. De Vault, et al., "Broad frequency shift of parametric processes in epsilon-near-zero time-varying media," in *Appl. Sci.* 10, 1318, 2020.
24. Taravati, Sajjad, and George V. Eleftheriades. "Pure and linear frequency-conversion temporal metasurface." in *Physical Review Applied* 15.6, 064011, 2021.
25. Taravati, Sajjad, and George V. Eleftheriades. "Microwave space-time-modulated metasurfaces." in *ACS Photonics* 9.2, 305-318, 2022.
26. Yonatan Sharabi, Alex Dikopoltsev, Eran Lustig, Yaakov Lumer, and Mordechai Segev, "Spatiotemporal photonic crystals," in *Optica* 9, 585-592, 2022.
27. A. A. Althuwayb *et al*., "Metasurface-Inspired Flexible Wearable MIMO Antenna Array for Wireless Body Area Network Applications and Biomedical Telemetry Devices," in *IEEE Access*, vol. 11, pp. 1039-1056, 2023.
28. Zainab S. Muqdad, Mohammad Alibakhshikenari, Taha A. Elwi, Zaid A. Abdul Hassain, Bal.S. Virdee, Richa Sharma, Salahuddin Khan, Nurhan Türker Tokan, Patrizia Livreri, Francisco Falcone, Ernesto Limiti," Photonic controlled metasurface for intelligent antenna beam steering applications including 6G mobile communication systems," in *AEU - International Journal of Electronics and Communications*, Volume 166, 2023, 154652, ISSN 1434-8411.
29. M. Alibakhshikenari *et al*., "A Comprehensive Survey on Antennas On-Chip Based on Metamaterial, Metasurface, and Substrate Integrated Waveguide Principles for Millimeter-Waves and Terahertz Integrated Circuits and Systems," in *IEEE Access*, vol. 10, pp. 3668-3692, 2022.
30. Alibakhshikenari, M., Virdee, B.S., Rajaguru, R.K. *et al.*, "High performance antenna-on-chip inspired by SIW and metasurface technologies for THz band operation," in *Sci Rep* 13, 56, 2023.
31. Wang, Xuchen, et al. "Metasurface-based realization of photonic time crystals." in *Sci Adv.* 9,14, 2023.
32. Naoto Tsuji, "Floquet States," in *arXiv* 2023, arXiv:2301.12676.
33. L. Brillouin, "Free electrons in metals and the role of Bragg reflections," in *J.Phys. Radium* 1, EDP Sciences: 377-400, 1930.

34. Charles Kittel, "Introduction to Solid State Physics," Wiley, 1996.
35. R. El-Ganainy, K. G. Makris, D. N. Christodoulides, et al., “Theory of coupled optical PT-symmetric structures,” in *Opt. Lett.* 32, 2632–2634, 2007.
36. A. Shojaeifard, K.-K. Wong, K.-F. Tong, et al., “Mimo evolution beyond 5g through reconfigurable intelligent surfaces and fluid antenna systems,” in *Proc. IEEE* 110, 1244–1265, 2022.